\documentclass[aps,twocolumn,groupeaddress,showkeys]{revtex4-1}
\usepackage[utf8]{inputenc} 
\usepackage{amsmath}
\usepackage{mathtools}
\usepackage{textcomp}
\usepackage{braket}
\usepackage{graphicx}
\usepackage{pgfplots}
\usepackage{csquotes}
\usepackage{xcolor}
\usepackage{hhline}
\usepackage{amssymb}

\usepackage{dcolumn}
\usepackage{tabularx}
\usepackage[colorlinks=true,linkcolor=blue,citecolor=blue,urlcolor=blue]{hyperref}
\usepackage{longtable}
\usepackage{braket}
\usepackage{float}
\newcolumntype{C}{>{\centering\arraybackslash}X}
\DeclareUnicodeCharacter{2212}{-}
\pgfplotsset{compat=1.5}
\begin{document}
\title{Playing Quantum Monty Hall Game in a Quantum Computer}

\author{Souvik Paul}
\email{sp17ms070@iiserkol.ac.in}
\author{Bikash K. Behera}
\email{bkb18rs025@iiserkol.ac.in}
\author{Prasanta K. Panigrahi}
\email{pprasanta@iiserkol.ac.in}
\affiliation{Department of Physical Sciences,\\ Indian Institute of Science Education and Research Kolkata, Mohanpur 741246, West Bengal, India}

\begin{abstract}
Here, we present the quantum version of a very famous statistical decision problem, whose classical version is counter-intuitive to many. The Monty Hall game can be phrased as a two person game between Alice and Bob. In their pioneering work, Flitney and Abbott [Phys. Rev. A \textbf{65}, 062318 (2002)] showed that by using a maximally entangled system for Alice and Bob's choices, and using quantum strategies, Bob and Alice can win or lose depending on the strategy chosen by either of the players. Here we develop a new quantum algorithm with quantum circuits for playing the quantum Monty Hall game by a user. Our quantum algorithm uses the quantum principles of superposition and entanglement so that it can be efficiently played on a quantum computer. We present two schemes, one calculating the probability of winning or loss and the other determining whether a player (say Alice) wins or not. 
\end{abstract}


\maketitle

\section{Introduction}

In the recent past, much interest has been developed in the discipline of quantum information \cite{qmh_NielsenCUP2000,qmh_KhanCTP2010} that has led to the creation of quantum game theory \cite{qmh_MeyerPRL1999,qmh_KhanCTP2010}. Quantum game theory \cite{qmh_EisertPRL1999,qmh_MarinattoPLA2000,qmh_FlitneyJPA2005,qmh_CheonPSJ2008,qmh_RamzanJPA2008,qmh_IqbalPLA2008,qmh_Flitney1PRA2002,qmh_KhanCTP2010} has attracted a lot of attention during the last few years. The quantum Monty Hall problem \cite{qmh_KhanCTP2010} is an interesting example in this realm. The quantum game theory has been shown to be experimentally feasible through the application of a measurement-based protocol by Prevedel \emph{et al.} \cite{qmh_PrevedelNJP2007}. They realized a quantum version of the Prisoner's Dilemma game based on the entangled photonic cluster states \cite{qmh_KhanCTP2010}. There are many paradoxes and unsolved problems associated with quantum information \cite{qmh_NielsenCUP2000} and the study of quantum game theory is a useful tool to explore this area. Another motivation is that in the area of quantum communication, optimal quantum eavesdropping can be treated as a strategic game with the goal of extracting maximal information \cite{qmh_BrandtPQE1998,qmh_Flitney1PRA2002}. It has also been suggested that a quantum version of the Monty Hall problem may be of interest in the study of quantum strategies of quantum measurement \cite{qmh_LiPLA2001,qmh_Flitney1PRA2002}.

The classical Monty Hall problem has raised much interest because it is sharply counter-intuitive. From an informational point of view, it clearly illustrates the case where a null operation provides the information about the system. In the classical Monty Hall game \cite{qmh_SavantAS1991,qmh_GillmanAMM1992} the host Bob secretly selects one door of three behind which to place a prize. The player Alice picks a door. Bob then opens another door where there is no prize. Alice now has the option of sticking with his current selection or changing to the untouched door. Classically, the optimum strategy for Alice is to alter her choice of door and this, surprisingly, doubles her chance \cite{qmh_SavantAS1991} of winning the prize from $\frac{1}{3}$ to $\frac{2}{3}$.

\section{Quantum Monty Hall Game \label{qmh_Problem}}

A number of researchers have contributed towards the quantization of Monty Hall problem \cite{qmh_KhanCTP2010,qmh_ZanderABAS2006,qmh_D'ArianoQIC2002}. For the Monty Hall game where both participants can apply quantum strategies, it has been observed that maximal entanglement of the initial states results the same payoffs as compared to the the classical game. The game is called to be fair, when the host, Bob, has access to a quantum strategy while Alice does not. Even if Bob can adopt a strategy with an expected payoff of $\frac{1}{2}$ for each person, Alice can win all the time if she has access to a quantum strategy and Bob does not. Non-entangled initial states produce the payoffs as the classical case as expected. Under certain operations, it is also possible for Bob to win the game with payoff $1$ \cite{qmh_Flitney1PRA2002}. Similar results have been recreated by D'Ariano \emph{et al.} \cite{qmh_D'ArianoQIC2002} by using density matrices and the concept of quantum notepads. Kurzyk and Glos \cite{qmh_KurzykQIP2016} have used an entangled state between Alice's first choice and the position of the prize. Using Bayesian networks, they have shown that the entanglement of quantum states has influence on the results of reasoning, and that there exists a quantum state for which the Monty Hall game is fair under the assumption that Bob and Alice do not have access to quantum strategies. Our present purpose here is to advance a different quantum version of the Monty Hall problem and to develop a quantum circuit \cite{qmh_ZanderABAS2006} so that it can be designed on a quantum computer and any user can play it. 

\section{Scheme-1: Determining the probability of winning or losing \label{qmh_Scheme1}} 
\begin{figure*}[]
\centering
\includegraphics[scale=0.62]{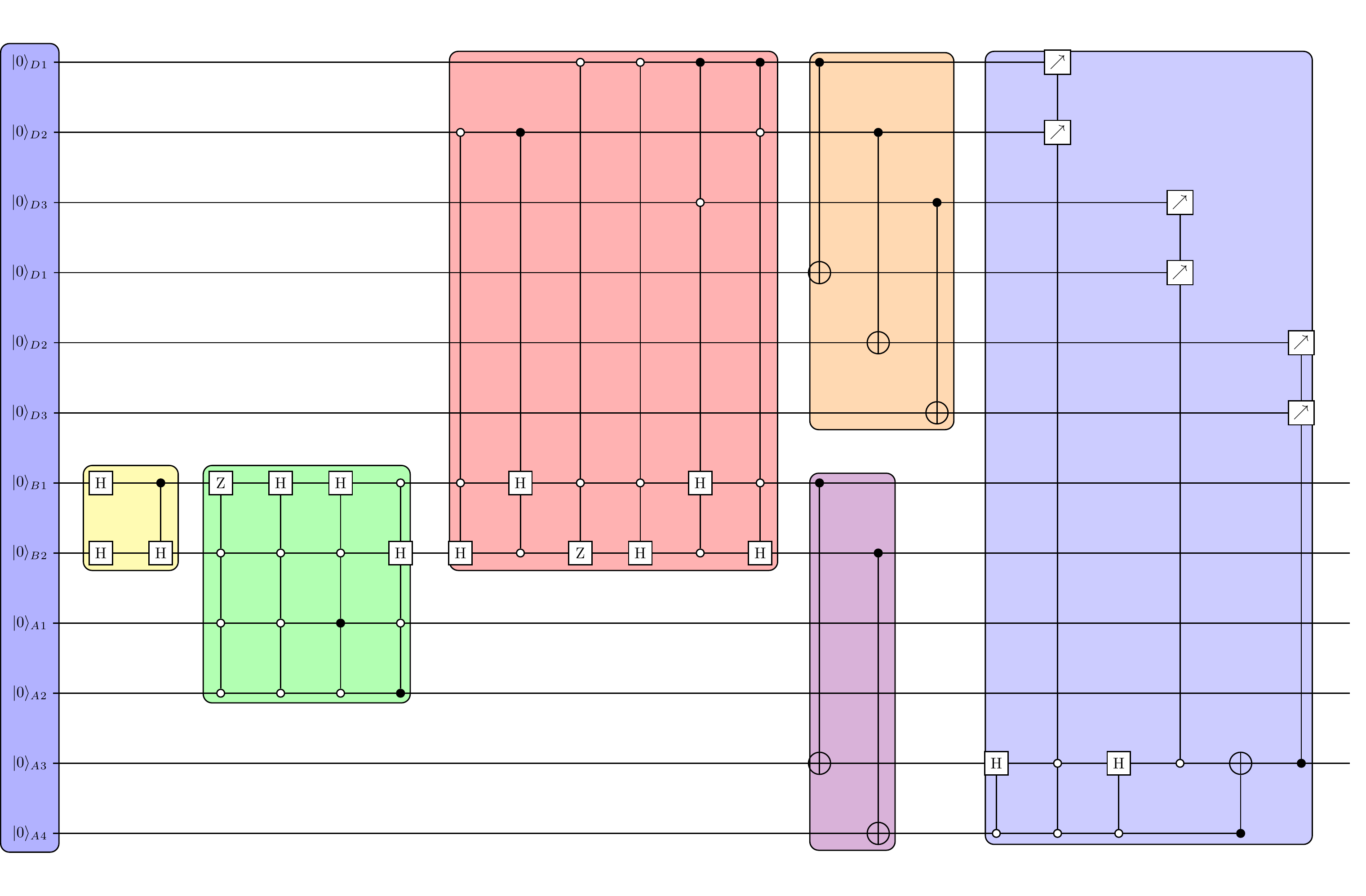}
\caption{\textbf{Quantum circuit illustrating Scheme-I}. \emph{The quantum circuit for the implementation of the Monty Hall game. A superposition of the initial states is created. Alice then chooses one of the three doors, accordingly Bob opens one of the three doors, depending on which door Alice chooses to open and the prize in behind which door. Alice then makes her second choice, and Bob opens the doors.}}
\label{qmh_Fig1}
\end{figure*}

Fig. \ref{qmh_Fig1} presents the quantum circuit explaining the scheme for playing the quantum Monty Hall game. As illustrated in the circuit, we take three qubits ($\Ket{0}_{D1}\Ket{0}_{D2}\Ket{0}_{D3}$) representing the three doors D1, D2, and D3. If there is a prize is in any of the door, then the corresponding qubit's state becomes $|1\rangle$. Three more qubits are taken for representing the doors D1, D2, and D3 as at the end of the quantum circuit, three controlled measurements needs to performed. Then we take two qubits ($\Ket{0}_{B1}$ and $\Ket{0}_{B2}$) for representing Bob's state for storing information about the doors and which doors open or not. Two qubits ($\Ket{0}_{A1}$ and $\Ket{0}_{A2}$) are used to Alice's two-qubit input. Then the two qubits ($\Ket{0}_{A3}$ and $\Ket{0}_{A4}$) are used to represent Alice's second choice which decides to which two doors to open.

The following initial assumptions are taken to proceed; 
\begin{itemize}
\item Door D1 is denoted by $\Ket{00}$
\item Door D2 is denoted by $\Ket{01}$
\item Door D3 is denoted by $\Ket{10}$
\end{itemize}

\textbf{Step-1:}  Before the game starts, Bob has classical information with regards to the door, behind which the prize is present. The first three qubits in the quantum circuit represents the three doors D1, D2, and D3 that are initialized in state $|0\rangle$. If the prize is present in any of the doors then the state of the corresponding qubit changes to $|1\rangle$ state. We denote the Bob's state by the qubits $B1$ and $B2$ (Fig. \ref{qmh_Fig1}). 

\textbf{Step-2:} Applying two Hadamard gates ($H$) on Bob's two qubits ($|00\rangle_{B1B2}$) each prepared in $|0\rangle$ state, we create a superposition of $|00\rangle$, $|01\rangle$, $|10\rangle$ and $|11\rangle$ states. Since we need superposition of only three states for representing the three doors' states, we remove $\Ket{11}$ state by using a controlled-Hadamard gate ($CH_{B1B2}$), where $B1$ qubit acts as the control and $B2$ qubit as the target.

\textbf{Step-3:} Now Alice chooses one of the three doors. Her choice can be any one from the following cases:

\begin{itemize}
\item If Alice chooses D1 i.e. $\Ket{00}$, then $\Ket{00}$ state needs to be removed from Bob's state, which can be achieved by using anti-controlled-Z ($ACZ_{B1B2}$) and anti-controlled-Hadamard ($ACH_{B1B2}$) gates. Now, the states with Bob are superposition of $\Ket{01}$ and $\Ket{10}$.
\item If Alice chooses D2, i.e. $\Ket{01}$, an anti-controlled-Hadamard ($ACH_{B1B2}$) will be applied on the Bob's qubit resulting a superposition of ($\Ket{00}$ and $\Ket{10}$) states.
\item If Alice chooses D3, i.e. $\Ket{10}$, Bob's state needs to be in superposition of $\Ket{00}$ and $\Ket{01}$ states, which is achieved by applying anti-controlled-Hadamard ($H_{B2B1}$) gate. 
\end{itemize}

\textbf{Step-4:} 
Here in this step, we make the following assumption, i.e., whenever Bob has the option of removing more than one states (i.e, opening more than one doors), he chooses to remove the state, i.e., opens the lowest door available to him (D1$<$D2$<$D3). According to the rule of the game, at first attempt of Alice, Bob would not open the door as asked by her, and also would open the door where there is no prize. However, in the second attempt of Alice, Bob has to follow her and exactly open the door as asked by her. According to the above rule, the following three cases can be considered depending upon the presence of prize in any of the doors; 

\textbf{Case-4A: Alice chooses D1:}
\begin{itemize}
\item The prize is in D3 (the state of the first three qubits ($\Ket{D1D2D3}=\Ket{001}$), then Bob opens the door D2 i.e., removes $\Ket{01}$ state using anti-controlled-Hadamard ($ACH_{B1B2}$) operation and Alice is left with a superposition state ($\Ket{A3A4}$) of $\Ket{00}$ and $\Ket{10}$.
\item The prize is in D2 ($\Ket{010}$), then Bob opens the door D3, i.e., removes the $\Ket{10}$ state using anti-controlled-Hadamard $ACH_{B2B1}$. Hence Alice is left with superposition of $\Ket{00}$ and $\Ket{01}$ states, i.e, $\Ket{A3A4}=\frac{\Ket{00}+\Ket{01}}{\sqrt{2}}$.
\item The prize is in D1 ($\Ket{010}$), then Bob opens D2 i.e., removes the $\Ket{01}$ state using an anti-controlled-Hadamard $H_{B1B2}$ operation. Thus Alice's state becomes, $\Ket{A3A4}=\frac{\Ket{00}+\Ket{10}}{\sqrt{2}}$. 
\end{itemize}

\textbf{Case-4B: Alice chooses D2:}
\begin{itemize}
\item The prize is in D3 ($\Ket{001}$), then Bob opens D1 i.e., $\Ket{00}$ is removed using anti-controlled-Z ($ACZ_{B1B2}$) and anti-controlled-Hadamard ($ACH_{B1B2}$) operations and Alice is left with a superposition state of $\Ket{01}$ and $\Ket{10}$.
\item The prize is in D2 ($\Ket{010}$), then Bob opens D1 i.e., removes $\Ket{00}$ state using anti-controlled-Z gate ($ACZ_{B1B2}$) and anti-controlled-Hadamard ($ACH_{B1B2}$) operations and Alice has the superposition state of $\Ket{01}$ and $\Ket{10}$.
\item The prize is in D1 ($\Ket{010}$), then Bob opens the door D3 i.e., removes $\Ket{10}$ state using anti-controlled-Hadamard ($ACH_{B2B1}$) and  Alice's state remains in superposition of $\Ket{00}$ and $\Ket{01}$.
\end{itemize}

\textbf{Case-4C: Alice chooses D3:}
\begin{itemize}
\item The prize is in D3 ($\Ket{001}$), then Bob opens D1 i.e., $\Ket{00}$ is removed using anti-controlled-Z ($ACZ_{B1B2}$) and anti-controlled-Hadamard ($ACH_{B1B2}$) operations and Alice has the superposition of $\Ket{01}$ and $\Ket{10}$.
\item The prize is in D2 ($\Ket{010}$), then Bob opens D1, i.e., $\Ket{00}$ is removed using anti-controlled-Z ($ACZ_{B1B2}$) and anti-controlled-Hadamard ($ACH_{B1B2}$) operations and Alice has the superposition of $\Ket{01}$ and $\Ket{10}$.
\item The prize is in D1 ($\Ket{010}$), then Bob opens D2, i.e., $\Ket{01}$ is removed using anti-controlled-Hadamard ($ACH_{B1B2}$) and Alice has the superposition of $\Ket{00}$ and $\Ket{10}$.
\end{itemize}

\textbf{Step-5:} An entanglement is created between the two states of Bob and Alice, which allows restricted information to be communicated to Alice (Bob communicates to Alice the state that has to be removed).

\textbf{Step-6:} The three super positions left with Alice are:
\begin{itemize}
\item $\Ket{00}$ and $\Ket{10}$ - 2nd measurement is restricted as Bob has already removed $\Ket{01}$ state (i.e, has opened the door D2).
\item $\Ket{01}$ and $\Ket{10}$ - 1st measurement is restricted as Bob has already removed $\Ket{00}$ state (i.e, has opened the door D1).
\item $\Ket{00}$ and $\Ket{01}$ - 3rd measurement is restricted as Bob has already removed $\Ket{10}$ state (i.e, has opened the door D3).
\end{itemize}
Hence controlled measurements are taken. The circuit is simulated and the probabilities of Alice's win are obtained. [See \textbf{Step 4}]

\section{Scheme-2: Determining Alice's win or loss}
\begin{figure*}[]
     \centering
    \includegraphics[scale=0.56]{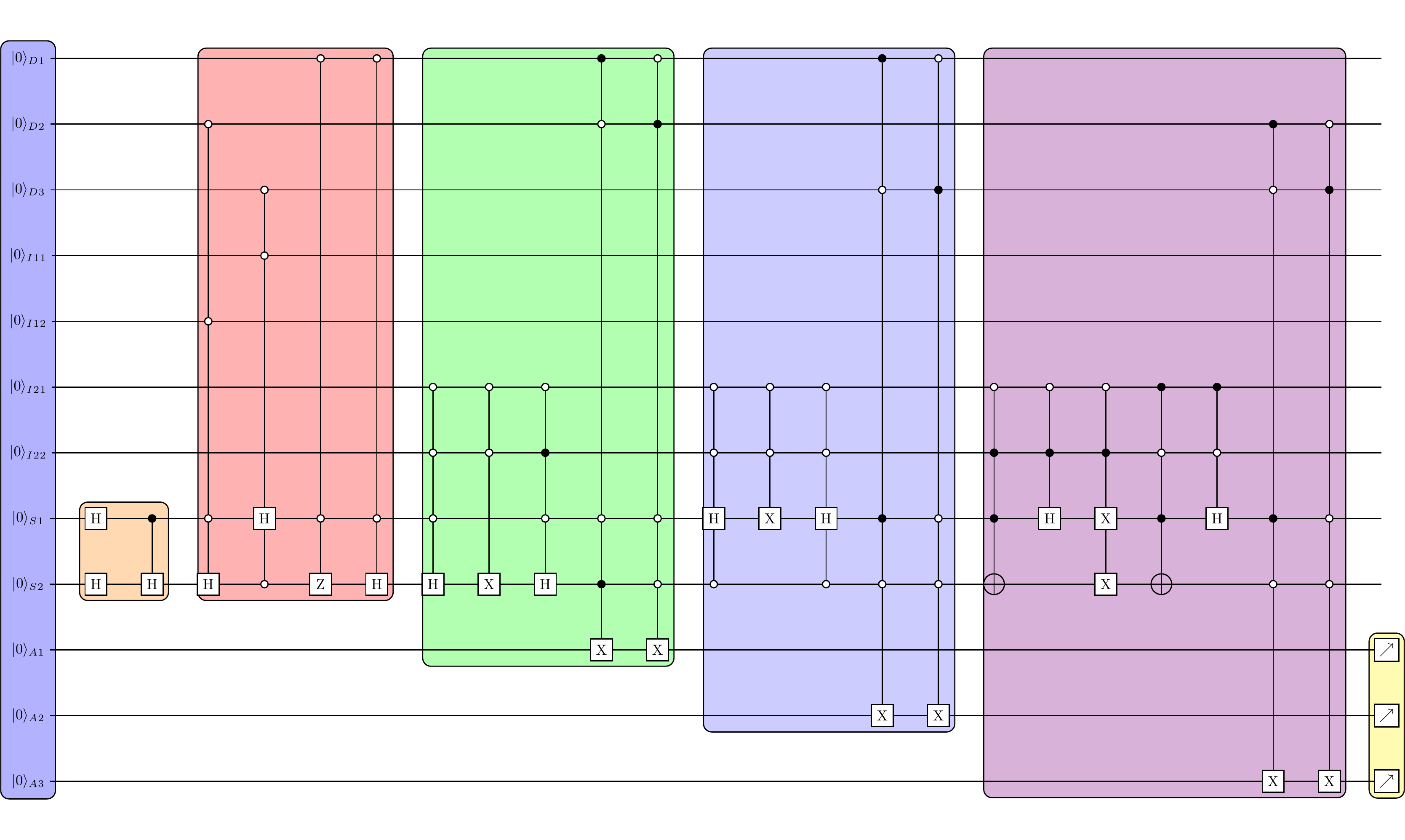}
    \caption{\textbf{Protocol - Scheme II}. \emph{The quantum circuit for the implementation of the Monty Hall game. A superposition of the initial states is done. Alice then chooses one of the three doors. Bob then opens one of the three doors, depending on which door Alice chooses to open and the prize in behind which door. Alice then makes her second choice. Bob opens the doors Alice chooses to open.}}
    \label{qmh_Fig2}
\end{figure*}

The quantum circuit provided in Fig. \ref{qmh_Fig2} depicts the winning or loss of the player, Alice. The details of the scheme is described in the following steps. As shown in the circuit, we take three qubits ($\Ket{0}_{D1}\Ket{0}_{D2}\Ket{0}_{D3}$) representing the three doors D1, D2, and D3 if a prize is there or not. For example, if the prize is in the second door D2, the three-qubit state becomes, $|010\rangle$, i.e., the presence of prize makes the qubit state to $|1\rangle$. Then four qubits, $\Ket{0}_{I11}$, $\Ket{0}_{I12}$, $\Ket{0}_{I21}$ and $\Ket{0}_{I22}$ are used for Alice's inputs, as she has to choose two doors one by one. Her possible inputs are $\Ket{00}$, $\Ket{01}$ and $\Ket{10}$. Then we take two qubits ($\Ket{0}_{S1}$ and $\Ket{0}_{S2}$) for creating a superposition of $\Ket{00}$, $\Ket{01}$, and $\Ket{10}$ states for representing the initial state when no doors are opened. Then three qubits ($\Ket{0}_{A1}$, $\Ket{0}_{A2}$ and $\Ket{0}_{A3}$) are assigned to Alice's qubits, which after the measurement at the end of the quantum circuit, determines whether Alice wins or loose.

The following initial assumptions are taken as the scheme-1 \ref{qmh_Scheme1}:
\begin{itemize}
\item Door D1 is denoted by $\Ket{00}$
\item Door D2 is denoted by $\Ket{01}$
\item Door D3 is denoted by $\Ket{10}$
\end{itemize}

\textbf{Step-1:}

We first create a superposition state is created to represent the initial state of the game when no doors are opened. This is done by using Hadamard gates ($H$) on the qubits ($\Ket{0}_{S1}$ and $\Ket{0}_{S2}$). Then applying a controlled-Hadamard ($CH_{S1S2}$) operation, a superposition of three states $\Ket{00}$, $\Ket{01}$ and $\Ket{10}$ is prepared.

\begin{figure*}[]
     \centering
    \includegraphics[scale=0.62]{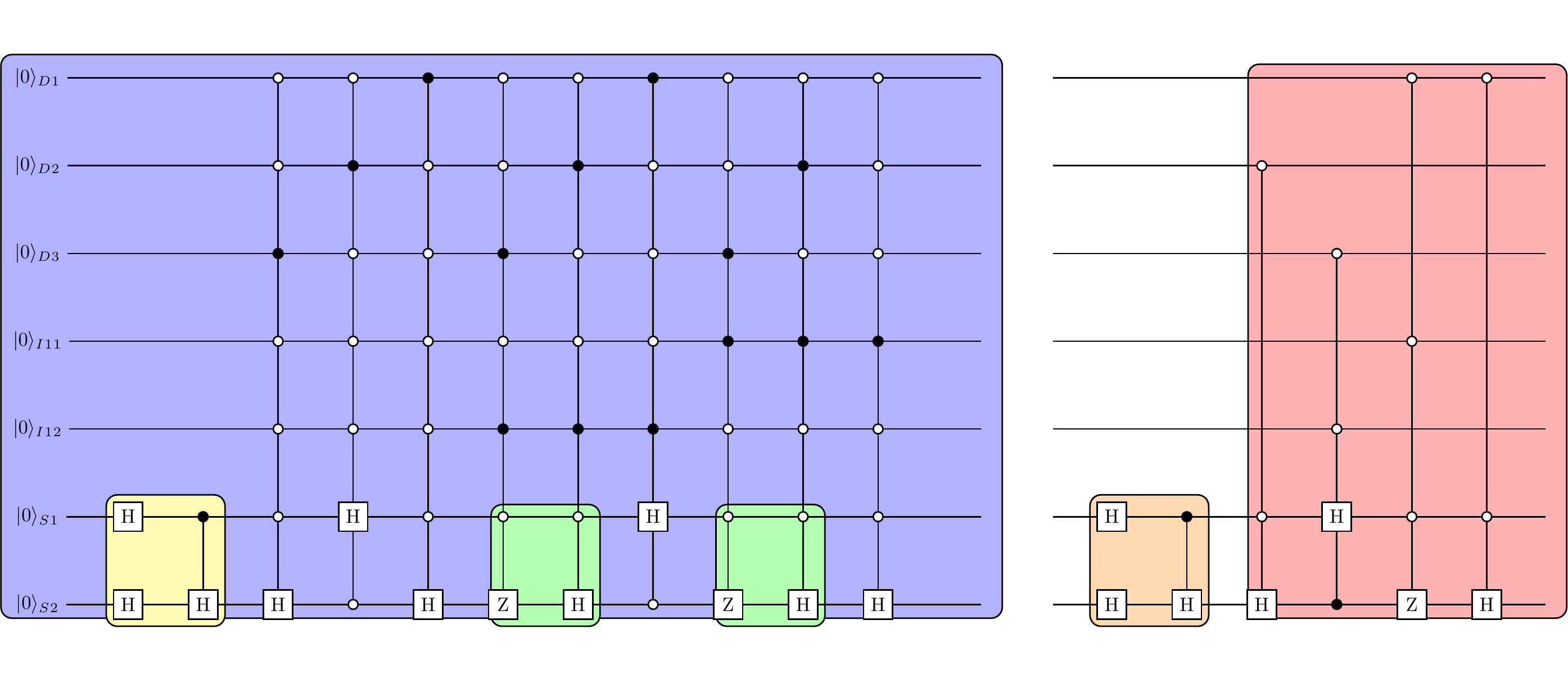}
    \caption{\textbf{Simplified Circuit in Scheme II}. \emph{The equivalent quantum circuit for creating a superposition state to represent the initial state of the game when no doors are opened and for obtaining the superposition of two states after Bob opens one of the doors, depending on where the prize is, Alice's choice of doors and our assumptions. All the 9 possible cases with 9 controlled operations are combined into 4 controlled operations as depicted in the right hand side circuit.}}
    \label{qmh_Fig3}
\end{figure*}

\textbf{Step-2:}
When the prize is in one of the three doors and Alice chooses to open one of the three doors, there are following 9 possible cases to be considered before one of the states is being removed from Bob's state. Here we make an assumption that whenever Bob has the option for opening one or more doors, then he chooses to open the door which is the lowest available to him (D1$<$D2$<$D3). According to the rule of the game, at first Bob would not open the door as asked by Alice, and also would not open the door where the prize is present. However, for the second time when asked by Alice, Bob has to open the door asked by her. Keeping in mind the above rule and conditions, the following 9 possible cases (Fig. \ref{qmh_Fig3}) arise before opening the first door.    

\textbf{Case-2A; Alice chooses D1 $\Ket{00}$:}
\begin{itemize}
    \item Prize is in D3 $\Ket{10}$. Bob opens D2 $\Ket{01}$ using anti-controlled-Hadamard ($ACH_{S1S2}$) and Alice has the superposition of $\Ket{00}$ and $\Ket{10}$.
    \item Prize is in D2 $\Ket{01}$. Bob opens D3 $\Ket{10}$ using anti-controlled-Hadamard, ($ACH_{S2S1}$) and Alice has superposition of $\Ket{00}$ and $\Ket{01}$.
    \item Prize is in D1 $\Ket{00}$. Bob opens D2 $\Ket{01}$ using anti-controlled-Hadamard, ($ACH_{S1S2}$) and Alice has superposition of $\Ket{00}$ and $\Ket{10}$.
\end{itemize}

\textbf{Case-2B; Alice chooses D2 $\Ket{01}$:}
\begin{itemize}
    \item Prize is in D3 $\Ket{10}$. Bob opens D1 $\Ket{00}$ using anti-controlled-Z gate ($ACZ_{S1S2}$) and anti-controlled-Hadamard ($ACH_{S1S2}$) and Alice has superposition of $\Ket{01}$ and $\Ket{10}$.
    \item Prize is in D2 $\Ket{01}$. Bob opens D1 $\Ket{00}$ using anti-controlled-Z gate ($ACZ_{S1S2}$) and anti-controlled-Hadamard, ($ACH_{S1S2}$) and Alice has the superposition of $\Ket{01}$ and $\Ket{10}$.
    \item Prize is in D1 $\Ket{00}$. Bob opens D3 $\Ket{10}$ using anti-controlled-Hadamard, ($ACH_{S2S1}$) and Alice has superposition of $\Ket{00}$ and $\Ket{01}$.
\end{itemize}

\textbf{Case-2C; Alice chooses D3 $\Ket{10}$:}
\begin{itemize}
    \item Prize is in D3 $\Ket{10}$. Bob opens D1 $\Ket{00}$ using anti-controlled-Z gate, ($ACZ_{S1S2}$) and anti-controlled-Hadamard, ($ACH_{S1S2}$) and Alice has superposition of $\Ket{01}$ and $\Ket{10}$.
    \item Prize is in D2 $\Ket{01}$. Bob opens D1 $\Ket{00}$ using anti-Controlled Z-gate, ($ACZ_{S1S2}$) and anti-controlled-Hadamard, ($ACH_{S1S2}$) and Alice has superposition of $\Ket{01}$ and $\Ket{10}$.
    \item Prize is in D1 $\Ket{00}$. Bob opens D2 $\Ket{01}$ using anti-controlled-Hadamard, ($ACH_{S1S2}$) and Alice has superposition of $\Ket{00}$ and $\Ket{10}$.
\end{itemize}

\textbf{Step-3:}

Now, for each superposition, there may be four cases. For example, in the superposition of $\Ket{00}$ and $\Ket{01}$, i.e. doors D1 and D2 remained to be opened, the possible cases are;
\begin{itemize}
    \item If prize is in D1, and Alice chooses D1, then she wins.
    \item If prize is in D1, and Alice chooses D2, then she loses.
    \item If prize is in D2, and Alice chooses D1, then she loses.
    \item If prize is in D2, and Alice chooses D2, then she wins.
\end{itemize}

For our scheme, we only consider the winning cases. In the cases Alice loses, the superposition states do not correspond to the state of the doors, and hence an Identity-I operation is performed on the ancilla state, resulting in no change in their states.

\textbf{Case-A; When the prize is either in D1 $\Ket{00}$ or D2 $\Ket{01}$:}
\begin{itemize}
    \item When the prize is in D1 $\Ket{00}$ and Alice chooses D1, an anti-controlled-Hadamard ($ACH_{S1S2}$) is applied on the superposition state of $\Ket{00}$ and $\Ket{01}$, after which X gate is applied on the second qubit to get $\Ket{01}$ state. 
    \item When the prize is in D2 $\Ket{01}$ and Alice chooses D2, an anti-controlled-Hadamard ($ACH_{S1S2}$) is applied on the superposition state of $\Ket{00}$ and $\Ket{01}$ to get $\Ket{00}$ state. 
\end{itemize}
Then two four-qubit-controlled-Not operations (the circuit in green in Fig. \ref{qmh_Fig2}) are applied to the first ancilla qubit ($\Ket{0}_{A1}$). If the ancilla qubit after the measurement is changed to $|1\rangle$ state, then Alice wins. 

\textbf{Case-B; When the prize is either in D1 $\Ket{00}$ or D3 $\Ket{10}$:}
\begin{itemize}
    \item When the prize is in D1 $\Ket{00}$ and Alice chooses D1, an anti-controlled-Hadamard ($ACH_{S2S1}$) is applied on the superposition state of $\Ket{00}$ and $\Ket{01}$, after which X gate is applied on the first qubit to get $\Ket{10}$ state.
    \item When the prize is in D3 $\Ket{10}$ and Alice chooses D3, an anti-controlled-Hadamard ($ACH_{S2S1}$) is applied on the superposition state of $\Ket{00}$ and $\Ket{01}$ to get $\Ket{00}$ state. 
\end{itemize}
Then two four-qubit-controlled-Not operations (the circuit in blue in Fig. \ref{qmh_Fig2}) are applied to the second ancilla qubit ($\Ket{0}_{A2}$). If the ancilla qubit after the measurement is changed to $|1\rangle$ state, then Alice wins.

\textbf{Case-C; When the prize is either in D2 $\Ket{01}$ or D2 $\Ket{10}$:}
\begin{itemize}
    \item When the prize is in D2 $\Ket{01}$ and Alice chooses D2, a controlled-NOT gate (CNOT) is applied on the superposition state of $\Ket{01}$ and $\Ket{10}$, after which Hadamard gate ($H$) is applied on the first qubit to get $\Ket{01}$ state.
    \item When the prize is in D3 $\Ket{10}$ and Alice chooses D3, a controlled-NOT gate (CNOT) is applied on the superposition state of $\Ket{01}$ and $\Ket{10}$, after which Hadamard gate ($H$) is applied on the first qubit to get $\Ket{01}$, which is further acted on by X gates on both the qubits, to get $\Ket{10}$ state. 
\end{itemize}
Then two four-qubit-controlled-Not operations (the circuit in purple in Fig. \ref{qmh_Fig2}) are applied to the third ancilla qubit ($\Ket{0}_{A3}$). If the ancilla qubit after the measurement is changed to $|1\rangle$ state, then Alice wins.

After the measurement of all the three ancilla qubits, if the outcomes is any of the states out of $\Ket{100}$, $\Ket{010}$ and $\Ket{001}$ states, then Alice wins otherwise looses. 

\section{Conclusion \label{qmh_conclusion}}

To conclude, we have presented a new scheme for the quantum version of the Monty Hall problem. We wish to bring to light the main features in which our present quantum version of the Monty Hall problem differs from the other quantum versions already discussed in the literature. It has similarities with the work of Flitney and Abbott (2002) \cite{qmh_Flitney1PRA2002}, Khan \emph{et al.} (2010) \cite{qmh_KhanCTP2010} and Kurzyk and Glos (2016) \cite{qmh_KurzykQIP2016}. However our scheme is different in the way that we have created a maximally entanglement pair between Alice and Bob in the latter part of the circuit. While earlier versions of this quantum game is modelled on the basis of a composite quantum system consisting of three qutrits: one associated with the location of the prize, a second one corresponding to Alice's choice and a third one associated with Bob's choice. In our discussed protocol, we have assigned states to the doors, while Alice and Bob perform operations on the doors, depending on certain restrictions.

It is shown that a fair two-party zero-sum game can be carried out if a player is permitted to adopt quantum measurement strategy, while in classical situation, the other player can always win with high probability \cite{qmh_LiPLA2001}. Since we have been able to design a circuit for a qutrit state using qubits, we have paved the path for future work in designing algorithms for quantum games and implementing circuits for the same. Further we would like to work on implementing quantum circuits for other games like the Prisoner's dilemna and other Bayesian games. Also, any two player game can be extended to three-player games and further to n-player games in general. Quantum games can be used to understand various other interesting systems like Optical networks, besides others \cite{qmh_RamosarXiv2005}. We are also interested in the experimental realization of the quantum Monty Hall problem and other games \cite{qmh_SchuckEQEC2003}. We also wish to increase the efficiency of our circuit by working on error correction and better algorithms \cite{qmh_Behera1QIP2018,qmh_SingharXiv2018,qmh_Behara2arXiv2018,qmh_Behera3arXiv2018,qmh_Behera4arXiv2017,qmh_Behera5arXiv2018,qmh_Behera6arXiv2018,qmh_Behera7arXiv2018,qmh_Behera8arXiv2018,qmh_Behera9arXiv2018,qmh_Behera10arXiv2018,qmh_Behera11arXiv2018,qmh_Behera12arXiv2018}. Studying quantum games and quantum problems like the Monty Hall problem motivates to develop advanced techniques more suited to contemporary practical problems.

\section{Acknowledgments}
\label{qlock_acknowledgments}
S.~P. would like to acknowledge the contributions of N.~N. Hegade, K. Haldar and R.~K. Singh in the undertaking of this paper. B.K.B. acknowledges the support of Institute fellowship provided by IISER Kolkata. The authors acknowledge the support of IBM Quantum Experience for producing experimental results. The views expressed are those of the authors and do not reflect the official policy or position of IBM or the IBM Quantum Experience team. 




\begin{thebibliography}{40}
\makeatletter
\providecommand \@ifxundefined [1]{%
 \@ifx{#1\undefined}
}%
\providecommand \@ifnum [1]{%
 \ifnum #1\expandafter \@firstoftwo
 \else \expandafter \@secondoftwo
 \fi
}%
\providecommand \@ifx [1]{%
 \ifx #1\expandafter \@firstoftwo
 \else \expandafter \@secondoftwo
 \fi
}%
\providecommand \natexlab [1]{#1}%
\providecommand \enquote  [1]{``#1''}%
\providecommand \bibnamefont  [1]{#1}%
\providecommand \bibfnamefont [1]{#1}%
\providecommand \citenamefont [1]{#1}%
\providecommand \href@noop [0]{\@secondoftwo}%
\providecommand \href [0]{\begingroup \@sanitize@url \@href}%
\providecommand \@href[1]{\@@startlink{#1}\@@href}%
\providecommand \@@href[1]{\endgroup#1\@@endlink}%
\providecommand \@sanitize@url [0]{\catcode `\\12\catcode `\$12\catcode
  `\&12\catcode `\#12\catcode `\^12\catcode `\_12\catcode `\%12\relax}%
\providecommand \@@startlink[1]{}%
\providecommand \@@endlink[0]{}%
\providecommand \url  [0]{\begingroup\@sanitize@url \@url }%
\providecommand \@url [1]{\endgroup\@href {#1}{\urlprefix }}%
\providecommand \urlprefix  [0]{URL }%
\providecommand \Eprint [0]{\href }%
\providecommand \doibase [0]{http://dx.doi.org/}%
\providecommand \selectlanguage [0]{\@gobble}%
\providecommand \bibinfo  [0]{\@secondoftwo}%
\providecommand \bibfield  [0]{\@secondoftwo}%
\providecommand \translation [1]{[#1]}%
\providecommand \BibitemOpen [0]{}%
\providecommand \bibitemStop [0]{}%
\providecommand \bibitemNoStop [0]{.\EOS\space}%
\providecommand \EOS [0]{\spacefactor3000\relax}%
\providecommand \BibitemShut  [1]{\csname bibitem#1\endcsname}%
\let\auto@bib@innerbib\@empty

\bibitem{qmh_NielsenCUP2000}M.~A. Nielson and I.~L. Chuang, Quantum Computation and Quantum Information, Cambridge University Press, Cambridge (2000).
\bibitem{qmh_KhanCTP2010}S. Khan, M. Ramzan, and M.~K. Khan, \href{\doibase 10.1088/0253-6102/54/1/10/pdf}{Commun. Theor. Phys. \textbf{54}, 47 (2010)}.
\bibitem{qmh_MeyerPRL1999}D.~A. Meyer, \href{\doibase 10.1103/PhysRevLett.82.1052}{Phys. Rev. Lett. \textbf{82}, 1052 (1999)}.
\bibitem{qmh_EisertPRL1999}J. Eisert, M. Wilkens, and M. Lewenstein, \href{\doibase 10.1103/PhysRevLett.83.3077}{Phys. Rev. Lett. \textbf{83}, 3077 (1999)}. 
\bibitem{qmh_MarinattoPLA2000}L. Marinatto and T. Weber, \href{\doibase 10.1016/S0375-9601(00)00441-2}{Phys. Lett. A \textbf{272}, 291 (2000)}.  
\bibitem{qmh_FlitneyJPA2005}A.~P. Flitney and D. Abbott, \href{\doibase 10.1088/0305-4470/38/2/011}{J. Phys. A \textbf{38}, 449 (2005)}.
\bibitem{qmh_CheonPSJ2008}T. Cheon and I. Iqbal, \href{\doibase 10.1143/JPSJ.77.024801}{Phys. Soc. Japan \textbf{77}, 024801 (2008)}. 
\bibitem{qmh_RamzanJPA2008}M. Ramzan, A. Nawaz, A.~H. Toor, and M.~K. Khan, \href{\doibase 10.1088/1751-8113/41/5/055307}{J. Phys. A.:Math. Theor. \textbf{41}, 055307 (2008)}. 
\bibitem{qmh_IqbalPLA2008}A. Iqbal, T. Cheon, and D. Abbott, \href{\doibase 10.1016/j.physleta.2008.09.026}{Phys. Lett. A \textbf{372}, 6564 (2008)}.
\bibitem{qmh_Flitney1PRA2002}A.~P. Flitney and D. Abbott, \href{\doibase 10.1103/PhysRevA.65.062318}{Phys. Rev. A \textbf{65}, 062318 (2002)}.
\bibitem{qmh_PrevedelNJP2007}R. Prevedel, A. Stefanov, P. Walther, and Z. Zeilinger, \href{\doibase 10.1088/1367-2630/9/6/205}{New J. Phys. \textbf{9}, 205 (2007)}. 
\bibitem{qmh_BrandtPQE1998}H.~E. Brandt, \href{\doibase 10.1016/S0079-6727(99)00003-8}{Prog. Quantum Electron. \textbf{22}, 257 (1998)}.
\bibitem{qmh_LiPLA2001}C.-F. Li, Y.-S. Zhang, Y.-F. Huang, and G.-C. Guo, \href{\doibase 10.1016/S0375-9601(01)00072-X}{Phys. Lett. A \textbf{280}, 257 (2001)}. 
\bibitem{qmh_SavantAS1991}M. vos Savant, Am. Stat \textbf{45}, 347 (1991). 
\bibitem{qmh_GillmanAMM1992}L. Gillman, Am. Math. Monthly \textbf{99}, 3 (1992).
\bibitem{qmh_ZanderABAS2006}C. Zander, M. Casas, A. Plastino, and A.~R. Plastino, \href{\doibase 10.1590/S0001-37652006000300003}{An. Braz. Acad. Sci. \textbf{78}, 417 (2006)}.
\bibitem{qmh_D'ArianoQIC2002}G.~M. D'Ariano, R.~D. Gill, M. Keyl, R.~F. Werner, B. Kummerer, and H. Maassen, \href{https://arxiv.org/abs/quant-ph/0202120}{Quantum Inf. Comput. \textbf{2}, 355 (2002)}.
\bibitem{qmh_KurzykQIP2016}D. Kurzyk and A. Glos, \href{\doibase 10.1007/s11128-016-1431-8}{Quantum Inf. Process. \textbf{15}, 4927 (2016)}.
\bibitem{qmh_BarencoPRA1995}A. Barenco, C.~H. Bennett, R. Cleve, N. Margolus, P. Shor, T. Sleator, J. Smolin, and H. Weinfurter, \href{\doibase 10.1103/PhysRevA.52.3457}{Phys. Rev. A \textbf{52}, 3457 (1995)}.
\bibitem{qmh_RamosarXiv2005}P.~B.~M. de Sousa and R.~V. Ramos, \href{https://arxiv.org/abs/1105.2289}{arXiv:1105.2289}. 
\bibitem{qmh_SchuckEQEC2003}C. Schuck, O. Schulz, C. Kurtseifer, H. Wienfurter, \href{\doibase 10.1109/EQEC.2003.1314280}{Euro. Quantum Elect. Conf. (2003)}.
\bibitem{qmh_Behera1QIP2018}D. Ghosh, P. Agarwal, P. Pandey, B.~K. Behera, and P.~K. Panigrahi, \href{\doibase 10.1007/s11128-018-1920-z}{Quantum Inf. Process. \textbf{17}, 53 (2018)}. 
\bibitem{qmh_SingharXiv2018}R.~K. Singh, B. Panda, B.~K. Behera, and P.~K. Panigrahi, \href{https://arxiv.org/abs/1807.02883}{arXiv:1807.02883}.
\bibitem{qmh_Behara2arXiv2018}D. Aggarwal,  S. Raj, B.~K. Behera, and P.~K. Panigrahi, \href{https://arxiv.org/abs/1806.00781}{arXiv:1806.00781}
\bibitem{qmh_Behera3arXiv2018}D. Joy, M. Sabir, B.~K. Behara, and P.~K. Panigrahi, \href{https://arxiv.org/abs/1807.03219}{	arXiv:1807.03219}.
\bibitem{qmh_Behera4arXiv2017}N.~K. Hegade, B.~K. Behera, and P.K. Panigrahi, \href{https://arxiv.org/abs/1712.07326}{arXiv:1712.07326}.
\bibitem{qmh_Behera5arXiv2018}M. Kapil, B.~K. Behera, and P.~K. Panigrahi, \href{https://arxiv.org/abs/1807.00521}{arXiv:1807.00521}.
\bibitem{qmh_Behera6arXiv2018}Y. Mohanta, D.~S. Abhishikth, K. Pruthvi, V. Kumar, B.~K. Behera, and P.~K. Panigrahi, \href{https://arxiv.org/abs/1807.00323}{arXiv:1807.00323}.
\bibitem{qmh_Behera7arXiv2018}Manabputra, B.~K. Behera, and P.K. Panigrahi, \href{https://arxiv.org/abs/1806.10229}{arXiv:1806.10229}.
\bibitem{qmh_Behera8arXiv2018}R. Jha, D. Das, A. Dash, S. Jayaraman, B.~K. Behera, and P.~K. Panigrahi, \href{https://arxiv.org/abs/1806.10221}{arXiv:1806.10221}.
\bibitem{qmh_Behera9arXiv2018}K. Srinivasan, S. Satyajit, B.~K. Behera, and P.K. Panigrahi, \href{https://arxiv.org/abs/1805.10928}{arXiv:1805.10928}.
\bibitem{qmh_Behera10arXiv2018}A. Dash, D. Sarmah, B.~K. Sarmah, and P.~K. Panigrahi, \href{https://arxiv.org/abs/1805.10478}{arXiv:1805.10478}.
\bibitem{qmh_Behera11arXiv2018}B.~K. Behera, T. Reza, A. Gupta, and P.~K. Panigrahi, \href{https://arxiv.org/abs/1803.06530}{arXiv:1803.06530}.
\bibitem{qmh_Behera12arXiv2018}K. Srinivasan, B.~K. Behera, and P.~K. Panigrahi, \href{https://arxiv.org/abs/1801.00778}{arXiv:1801.00778}.

\end{thebibliography}
\end{document}